\documentstyle{article}

\begin{document}

\title{\bf Improved Phenomenological Renormalization Schemes}

\author{M.~A.~Yurishchev}
\medskip
\date{\sl
Vasilsursk Laboratory,
Radiophysical Research Institute,
606263 Vasilsursk,
Nizhny Novgorod Region,
Russia}

\maketitle

\begin{abstract}
An analysis is made of various methods of phenomenological renormalization
based on finite-size scaling equations for inverse correlation
lengths, the singular part of the free energy density, and their
derivatives. The analysis is made using two-dimensional Ising and Potts
lattices and the three-dimensional Ising model. Variants of equations for
the phenomenological renormalization group are obtained which ensure more
rapid convergence than the conventionally used Nightingale phenomenological
renormalization scheme. An estimate is obtained for the critical
finite-size scaling amplitude of the internal energy in the
three-dimensional Ising model. It is shown that the two-dimensional Ising
and Potts models contain no finite-size corrections to the internal
energy so that the positions of the critical points for these models can
be determined exactly from solutions for strips of finite width. It is
also found that for the two-dimensional Ising model the scaling
finite-size equation for the derivative of the inverse correlation
length with respect to temperature gives the exact value of the thermal
critical exponent.
\end{abstract}



\newpage

\section{Introduction}
\label{sec:Intro}

As we know \cite{M80}, the renormalization group method
lies at the basis of modern phase transition theory. In
this approach, the critical points of a system are identified
with the fixed points of renormalization group
mappings. By linearizing the renormalization group
transformations near a fixed point it is possible to find
the critical exponents of the system of interest.

Many different specific realizations of the renormalization
group procedure now exist. One of these realizations is
the so-called phenomenological renormalization
group proposed by Nightingale \cite{Nigh76,Nigh79} (see also the
reviews \cite{Nigh82,Nigh90}).
In this approach the scaling functional
relationship between the correlation lengths of partially
finite subsystems is interpreted as a renormalization
group transformation. Combining this with expressions
for the correlation lengths in terms of eigenvalues of the
transfer matrices reduces the initial functional equation
to a transcendental one and its solution gives an estimate
for the critical temperature.

Dos Santos and Sneddon \cite{DSS81} noted that a
phenomenological renormalization group can be constructed
using not only the correlation length but any quantity
having a power divergence at the phase transition point.

Binder \cite{B81a,B81b} suggested a variant of the
phenomenological renormalization group based on using the
moments of the distribution function of the order parameter;
the moments can be expressed directly in terms of the
susceptibility of the system. This method has been widely
used in calculations using the Monte Carlo method.
Recently, Itakura \cite{I96} expanded the Binder
phenomenological renormalization scheme using linear
combinations of different-order moments.

In the present study various phenomenological
renormalization variants using a wide range of physical
quantities (such as the free and internal energy, ordinary
and nonlinear susceptibility, and so on) are used to
calculate the critical coupling in the two-dimensional
Ising and Potts models and also in the three-dimensional
Ising model. An analysis can reveal the best
renormalization strategies. Increasing the convergence
is fundamentally important for three-dimensional systems
because as a result of the giant orders of the transfer
matrices, the eigenvalue problem can only be solved
for subsystems having extremely small transverse
dimensions.

However, a study of the convergence of various types
of renormalizations revealed equations which can be used
to find exact values of the parameters of infinite systems
(critical temperatures and in one case, critical exponent) using
the characteristics of partially finite subsystems.

This article is constructed as follows. Section 2 contains
equations for different variants of the phenomenological
renormalization group. In Section 3 these equations are
applied to the two-dimensional Ising and Potts
models and in Section 4 they are applied to the three-dimensional
Ising lattice. The conclusions reached are
summarized briefly in the final section.


\section{Equations for the phenomenological renormalization group}
\label{sec:PRGE}

In accordance with finite-size scaling theory \cite{F70,FB72,B83,P90},
the inverse correlation length $\kappa_L$ and the
singular part of the reduced free energy density
$f_L^s$ ($f_L^s=f_L-f_\infty$,
, where $f_L$ and $f_\infty$ are the free energy densities of
the subsystem and the total system, respectively) near
the critical point $t=h=0$ ($t=K-K_c$ is the "temperature" or
more accurately the deviation from the critical
coupling, and $h$ is the external field) satisfy the functional
relationships
\begin{equation}
   \label{eq:kappa}
   \kappa_L(t, h)
   = b^{-1}\kappa_{L/b}(t', h'),
\end{equation}
\begin{equation}
   \label{eq:f^s}
   f_L^{s}(t, h)
   = b^{-d}f_{L/b}^{s}(t', h').
\end{equation}
Here $d$ is the space dimension, $L$ is the characteristic
linear size of the subsystem, $b = L/L'$ is the coefficient
of the scaling transformation. The renormalization group
equations $t'=t'(t, h)$ and $h'=h'(t, h)$ linearized near
the fixed point $t=h=0$ have the form
\begin{equation}
   \label{eq:t'}
   t' = b^{y_t}t + O(t^2),
\end{equation}
\begin{equation}
   \label{eq:h'}
   h' = b^{y_h}h + O(h^2),
\end{equation}
where $y_t$ and $y_h$ are the thermal and magnetic critical
exponents of the system.

In the Nightingale approach \cite{Nigh76,Nigh79,Nigh82,Nigh90}
the relationship (\ref{eq:kappa})
is treated as a renormalization group mapping
$(t,h)\to(t',h')$.
In this case, the inverse correlation length of the
spin-spin correlation function is given by the standard
expression
\begin{equation}
   \label{eq:kappa_L}
   \kappa_L=\ln (\lambda_1^{(L)}/|\lambda_2^{(L)}|),
\end{equation}
where $\lambda_1^{(L)}$ and $\lambda_2^{(L)}$ are the largest
and second largest eigenvalues of the transfer matrix of the
subsystem, respectively.
As a result, for systems with phase diagram symmetry $h\to-h$
(and thus $h_c=0$; we only consider these systems here) the
only unknown coordinate of the critical point can be determined
from the transcendental equation
\begin{equation}
   \label{eq:kappaKc}
   L\kappa_L(K_c)=(L-1)\kappa_{L-1}(K_c),
\end{equation}
(in order to maximize the accuracy of the estimate $K_c$
it is usually assumed that $L'=L-1$). Nightingale described
his renormalization method as "phenomenological" \cite{Nigh79}
(see also \cite{KS81}) since it uses a scaling finite-size
equation which is not derived microscopically using the method.

Quite clearly, the phenomenological renormalization
$(t,h)\to(t',h')$ can be made with equal justification using
relation (\ref{eq:f^s}).
In this case, we obtain the following equation to determine
the critical coupling $K_c$
\begin{equation}
   \label{eq:fKc}
   L^df^s_L(K_c)=(L-1)^df^s_{L-1}(K_c).
\end{equation}
The free energy density of the partially finite subsystem
$L^{d-1}\times\infty$ is then calculated using the formula
\begin{equation}
   \label{eq:fL}
   f_L=L^{1-d}\ln\lambda^{(L)}_1,
\end{equation}
and the "background" $f_\infty$ is taken from outside.

Differentiating expressions (1) and (2) $m$ times with
respect to temperature and $n$ times with respect to the
field allowing for (3) and (4), we obtain [15]
\begin{equation}
   \label{eq:kappa^mn}
   L^{1-my_t-ny_h}\kappa_L^{(m,n)}(t, h)
   = (L')^{1-my_t-ny_h}\kappa_{L'}^{(m,n)}(t', h'),
\end{equation}
\begin{equation}
   \label{eq:f^smn}
   L^{d-my_t-ny_h}f_L^{s\,(m,n)}(t, h)
   = (L')^{d-my_t-ny_h}f_{L'}^{s\,(m,n)}(t', h').
\end{equation}
These relationships or some combinations thereof can
also be considered as implicit renormalization group
transformations. For large subsystem sizes all
equations of this type should give the same results.
However, for small L which we need to deal with in
practice,  the various renormalization group equations
yield estimates of different accuracy. As we shall see, this
circumstance means that it is possible to construct
phenomenological renormalization schemes which give faster
convergence than that achieved in conventionally used
realizations of the phenomenological renormalization
group method.

In addition to equations (\ref{eq:kappaKc}) and (\ref{eq:fKc}),
in the present study we also give estimates of the critical coupling
obtained from the following renormalization group
equations [taking those ratios (\ref{eq:kappa^mn}) and
(\ref{eq:f^smn}), from which the unknown exponents $y_t$ and
$y_h$ are dropped {\em a priori\/}]:
\begin{equation}
   \label{eq:chi4}
   \frac{\chi_L^{(4)}}{L^d\chi_L^2}\Big|_{K_c}=
   \frac{\chi_{L-1}^{(4)}}{(L-1)^d\chi_{L-1}^2}\Big|_{K_c},
\end{equation}
where $\chi_L=\partial^2f_L/\partial h^2|_{h=0}=f_L^{s\,(0,2)}$
is the susceptibility and
$\chi_L^{(4)}=\partial^4f_L/\partial h^4|_{h=0} =f_L^{s\,(0,4)}$
is the nonlinear susceptibility (equation (\ref{eq:chi4})
corresponds to the Binder phenomenological renormalization
\cite{B81a,B81b});
\begin{equation}
   \label{eq:kappa1}
   L^{2-d}(\kappa_L^{(1)})^2/\chi_L=
   (L-1)^{2-d}(\kappa_{L-1}^{(1)})^2/\chi_{L-1};
\end{equation}
\begin{equation}
   \label{eq:kappa2}
   L^{1-d}\kappa_L^{(2)}/\chi_L=
   (L-1)^{1-d}\kappa_{L-1}^{(2)}/\chi_{L-1};
\end{equation}
\begin{equation}
   \label{eq:kappa4}
   L^{1-2d}\kappa_L^{(4)}/\chi_L^2=
   (L-1)^{1-2d}\kappa_{L-1}^{(4)}/\chi_{L-1}^2.
\end{equation}
Here we have
$\kappa_L^{(n)}=\partial^n\kappa_L/\partial
h^n|_{h=0}=\kappa_L^{(0,n)}(K,0)$.
Formulas for the derivatives of the inverse correlation length
and the free energy with respect to $h$, expressed in terms of
the eigenvalues and eigenvectors of the transfer matrices,
suitable for programming are given in \cite{Yu94,Yu97}.

For better clarity, equations
(\ref{eq:kappaKc}), (\ref{eq:fKc}), (\ref{eq:chi4}), (\ref{eq:kappa1}),
(\ref{eq:kappa2}), and (\ref{eq:kappa4})
will be denoted by the mnemonic symbols
``$\kappa$'', ``$f^s$'',
``$\chi^{(4)}/\chi^2$'', ``$(\kappa^{(1)})^2/\chi$,
``$\kappa^{(2)}/\chi$'', and ``$\kappa^{(4)}/\chi^2$'',
respectively when the numerical data are given in the tables.


\section{Two-dimensional systems}
\label{sec:2DI}

We shall first test various types of phenomenological
renormalization group equations on the two-dimensional Ising
and Potts models, for which the positions
of the critical points are known in exact analytical form
\cite{Bax85,W82}.

\subsection{Ising model}

We shall start with the most studied system, the
Ising model. Table~\ref{tab:2DI} gives results of our calculations of
the critical coupling for a simple square Ising lattice.
The calculations were made for $L\times\infty$ strips having a
periodic boundary condition in the transverse direction.
Estimates are given for $(L-1,L)$ pairs with $L\leq5$.
For a $(3,4)$ pair the relative errors of the estimates are also
given in parentheses; their sign indicates whether they
are lower (-) or upper (+) estimates. The types of renormalization
group equations used are indicated in the
first column. We recall that in this model we have
$$
K_c={1\over2}\ln (1+\sqrt2),\qquad
f_\infty=\frac{2G}{\pi} + {1\over2}\ln 2
$$
($G$ is the Catalan constant) [20].

It can be seen from the data presented in Table~\ref{tab:2DI}
that the best lower estimate of $K_c$ is obtained from equation
(\ref{eq:kappa4}).
The Binder scheme gives slightly inferior results.
This approach, which is usually used for Monte Carlo
simulations, as we noted in the Introduction [because it
does not have formulas of the type (\ref{eq:kappa_L}) for
the inverse correlation length], was used in the transfer matrix
variant in \cite{SD85}. The Nightingale renormalization (first row
in Table~\ref{tab:2DI}) which is almost universal (except perhaps
for the study just mentioned \cite{SD85}) is used for calculations
using the transfer matrix method and is only third
in terms of accuracy among the lower estimates.

It is worth noting that we also succeeded in obtaining
phenomenological renormalization group equations
which also yield upper estimates of $K_c$ (the two lower
rows in Table~\ref{tab:2DI}). Of these two estimates the
renormalization scheme using the singular part of the free energy
density [equation(\ref{eq:fKc})] gives better-quality results.
The values obtained using this scheme are at the same time
the best in terms of absolute value among all the lower
and upper estimates.

Unfortunately the renormalization procedure using
the singular part of the free energy requires an advance
knowledge of the critical free energy of the system $f_\infty$.
In order to overcome this constraint, instead of the two
strips required to construct the renormalization group
equation, we take three. In other words we shall take
the triads $(L_1,L_2,L_3)$; an additional relationship can
eliminate the parameter $f_\infty$.
For the free energy densities of the three subsystems at the
critical point we have
\begin{equation}
   \label{eq:f123}
   f_{L_1} = f_\infty + A_f/L_1^d,\quad
   f_{L_2} = f_\infty + A_f/L_2^d,\quad
   f_{L_3} = f_\infty + A_f/L_3^d.
\end{equation}
Here $A_f=L^df_L^s(0,0)$ is the critical finite-size
scaling amplitude of the singular part of the free energy
density. Eliminating the parameters $f_\infty$ and $A_f$ from
the system (\ref{eq:f123}), we arrive at the equation
\begin{equation}
   \label{eq:FF}
   {\cal F}(K) = F(K) -
   \frac{L_2^d - L_1^d}{L_3^d - L_2^d}
   \biggl(\frac{L_3}{L_1}\biggr)^d = 0,
\end{equation}
where
\begin{equation}
   \label{eq:F}
   F(K) = \frac{f_{L_1} - f_{L_2}}{f_{L_2} - f_{L_3}}.
\end{equation}
Equation (\ref{eq:FF}) can apparently be used to determine the
position of the critical point but it has no real roots.

All the eigenvalues of the transfer matrix of an $L\times\infty$
Ising cylinder are known in exact analytic form for any
$L$ \cite{O44,K49}. The largest eigenvalue is
\begin{equation}
   \label{eq:lam1}
   \lambda_1^{(L)} = (2\sinh 2K)^{L/2}\exp [(\gamma_1 +
   \gamma_3 + \ldots + \gamma_{2L-1})/2],
\end{equation}
where $\gamma_r$ are the positive solutions of the equations
\begin{equation}
   \label{eq:chgam}
   \cosh\gamma_r = \cosh 2K\coth 2K - \cos (\pi r/L).
\end{equation}
Using formula (\ref{eq:fL}) we then obtain expressions for the
free energy densities $f_{L_i}$ $(i=1,2,3)$ contained in
(\ref{eq:F}).

Subsequent calculations using these formulas show
that the curves ${\cal F}(K)$ do not intersect the abscissa and
thus the equation ${\cal F}(K)=0$ has no real solutions (see
Fig. 1). However, it is noticeable that for various choices
of parameters $L_1$, $L_2$, and $L_3$ the function ${\cal F}(K)$
[and therefore $F(K)$] has an extremum positioned at the exact
value of the critical coupling. Differentiating relation
(\ref{eq:F}) with respect to $K$, we arrive at the equation
\begin{equation}
   \label{eq:uf}
   (u_{L_1} - u_{L_2})(f_{L_2} - f_{L_3}) =
   (u_{L_2} - u_{L_3})(f_{L_1} - f_{L_2}),
\end{equation}
where $u_L=\partial f_L/\partial K$ is the internal energy of the
subsystem. By using this equation or finding the extremum
of the function $F(K)$ we can extract exact values of the
critical temperatures for infinite systems (at any rate for
some) from the characteristics of their finite subsystems.
(This is the desired aim of any finite-size theory.)

Equation (\ref{eq:uf}) can be approached from several positions.
We shall analyze expression (\ref{eq:f^smn}) in the first
derivative with respect to $K$ ($m=1$, $n=0$). For a fixed
point this renormalization group transformation gives
\begin{equation}
   \label{eq:usKc}
   L^{d-y_t}u_L^s(K_c)
   = (L')^{d-y_t}u_{L'}^s(K_c),
\end{equation}
where $u^s_L=u_L-u_\infty$ is the singular part of the internal
energy density. It follows from (\ref{eq:chgam}) that
\begin{equation}
   \label{eq:dgam}
   \partial\gamma_r/\partial K|_{K=K_c} = 0 \quad (r\neq0)
\end{equation}
(in this context see Fig.~118 in the book \cite{H66}).
After differentiating we find that $u_L(K_c)=\sqrt{2}$
regardless of $L$.
In other words, all the finite-size corrections to the
background $u_\infty$ ($=\sqrt{2}$ \cite{O44}) are zero,
i.e., the singular part of the internal energy is completely
absent. Consequently, we obtain
\begin{equation}
   \label{eq:usL}
   u^s_L(K_c) = 0.
\end{equation}
This property of the model leads to equations (\ref{eq:uf}) and
(\ref{eq:usKc}) being satisfied.

We shall now analyze the first derivative with respect
to temperature of the other fundamental equation in
finite-size scaling theory, i.e., equation (1).

The second eigenvalue of the transfer matrix of the
Ising model in terms of value is given by
\begin{equation}
   \label{eq:lam2}
   \lambda_2^{(L)} = (2\sinh 2K)^{L/2}\exp [(\gamma_0 +
   \gamma_2 + \ldots + \gamma_{2L-2})/2].
\end{equation}
Equations (5), (18), (19), and (24) can be used to find
an analytic expression for the inverse correlation
length. Taking the first derivative with respect to $K$ we
obtain the dependencies
$\dot\kappa_L(K)\equiv\partial\kappa_L/\partial K$ for
various $L$.
All these dependencies have a common point of self-intersection
and this is positioned at the exact value of
the critical coupling: $\dot\kappa_L(K_c)=-2$ for all $L=1,2,\ldots$
(i.e. in an infinite lattice $\dot\kappa_\infty=-2$).
Thus we have
\begin{equation}
   \label{eq:dkap}
   \dot\kappa_L(K_c)=\dot\kappa_{L'}(K_c).
\end{equation}
Reversing this reasoning, we conclude that equation
(\ref{eq:dkap}) can also be used to determine the exact
position of the critical point of an infinite two-dimensional
Ising lattice using the solutions for strips of finite width.

Moreover, in accordance with relation (\ref{eq:kappa^mn}) for
$m=1$ and $n=0$ we have
\begin{equation}
   \label{eq:yt}
   y_t = 1 + \frac{\ln (\dot\kappa_L/\dot\kappa_{L'})}{\ln (L/L')}.
\end{equation}
Consequently, this finite-size expression taking into
account condition (\ref{eq:dkap}) gives an exact value for
the thermal critical exponent of the two-dimensional Ising
model: $y_t=1$.

For the Ising model equation (\ref{eq:dkap}) can be generalized
to the form
\begin{equation}
   \label{eq:dkap123}
   (\dot\kappa_{L_1}-\dot\kappa_{L_2})(\kappa_{L_2}-\kappa_{L_3})
   =(\dot\kappa_{L_2}-\dot\kappa_{L_3})(\kappa_{L_1}-\kappa_{L_2}).
\end{equation}
and the problem of determining $K_c$ can be reduced to
searching for the extremum of the function
\begin{equation}
   \label{eq:Phi}
   \Phi=\frac{\kappa_{L_1}-\kappa_{L_2}}{\kappa_{L_2}-\kappa_{L_3}}.
\end{equation}

\subsection{Potts model}

In the two-dimensional q-state Potts model for
an isotropic square-cell lattice the critical coupling is
$K_c=\ln (1+\sqrt{q})$ \cite{P52}.
Table~\ref{tab:2DP} contains estimates of $K_c$
obtained by us using various constructions of the phenomenological
renormalization group for the Potts
model with three states ($q=3$). The free energy
background in this model is given by \cite{Bax73,Bax82}
\begin{equation}
   \label{eq:f8}
   f_\infty = 4G/3\pi + \ln (2\sqrt{3}) + {1\over3}\ln (2 + \sqrt{3}).
\end{equation}
An analysis of the data given in Table~\ref{tab:2DP} shows that
equations (\ref{eq:fKc}) and (\ref{eq:kappa1}) give substantially
better-quality estimates than the Nightingale approach (first row in
the table). The renormalization group equation based
on the scaling relationship for the singular part of the
free energy again gives the smallest errors in terms of
absolute value.

Calculations made for $L\times\infty$ strips with a periodic
boundary condition in the transverse direction then
uniquely indicate that in the two-dimensional q-state
Potts model all finite-size corrections to
the internal energy ($u_\infty=1+1/\sqrt{q}$ \cite{Bax85})
are also absent at the critical point. Consequently, the
equation
\begin{equation}
   \label{eq:uu}
   u_L(K_c) = u_{L'}(K_c)
\end{equation}
gives the exact value of $K_c$.
Satisfying condition (\ref{eq:uu})
then has the result that equation (\ref{eq:uf}) is valid and thus
the extremum of the function (\ref{eq:F}) is positioned strictly
at the exact value of the critical coupling. It should be
noted that in computer calculations it is more logical to
search directly for the extremum of the function than to
differentiate the free energy numerically and then solve
the transcendental equation (\ref{eq:uu}) numerically.

Equation (\ref{eq:uu}) for the pair $(1,2)$ was first obtained by
Wosiek from his postulated maximum principle for the
normalized moments of transfer matrices \cite{W94} (see also
\cite{BW94}). In \cite{P95} the Wosiek equation was then
generalized to the form (\ref{eq:uu}). Our equation (20) is a
further generalization of (\ref{eq:uu}).

Note that in the Potts model the curves $\dot\kappa_L(K)$ also
intersect. However, unlike the Ising model the points of
intersection are not now positioned at the exact value of
$K_c$.
At the exact value of the critical coupling the functions
$\dot\kappa_L(K)$ have different values for various $L$.
Unfortunately the scaling formula (\ref{eq:yt}) then only gives
approximate estimates of the thermal critical exponent
which, for example for the three-site Potts model is $y_t=6/5$.

Returning to the two-dimensional Ising model we can confirm that a
complete system of equations of the type (\ref{eq:uu}) is in fact
obtained for this model:
\begin{equation}
   \label{eq:uuj}
   u^j_L(K_c) = u^j_{L'}(K_c),
\end{equation}
where
$u^j_L=\partial f^j_L/\partial K$
are internal energy levels; this
term was taken by analogy with the "free energy levels"
$f^j_L= L^{1-d}\ln\lambda^{(L)}_j$, $j=1,2,\ldots$ (see,
for example [30]).


\section{Three-dimensional Ising lattice}

We shall now consider a three-dimensional Ising
model for a simple cubic lattice. Unlike the two-dimensional
Ising model, which is solved analytically over the
entire temperature range, or the two-dimensional Potts
model which is solved exactly at a single point, the critical
point, for the three-dimensional Ising model we are
forced to confine ourselves to numerical estimates. The
highest-quality estimates for the critical coupling and
the free energy background have now been obtained by
Monte Carlo simulation for $L\times L\times L$ cubes:
$K_c=0.22165459(10)$ \cite{BST99} and $f_\infty=0.777\,90(2)$
\cite{M89}.

Table~\ref{tab:3DI} gives our results for the three-dimensional
Ising model. The renormalizations were made for
$L\times L\times\infty$ parallelepipeds with periodic boundary
conditions in both transverse directions. The calculations
were made for subsystems with side lengths $L\leq4$, i.e.
for transfer matrices having dimensions up to
$65\,536\times65\,536$.

We can see from Table~\ref{tab:3DI} that as in the two-dimensional
case, the best lower estimates of $K_c$ are provided
by the renormalization group equation (\ref{eq:kappa4}).
On comparing the lower and upper estimates we can conclude
that the best of all the results is again given by renormalizing
the singular part of the free energy.

In the three-dimensional case the amplitudes of the
finite-size corrections to the internal energy at
the critical point are no longer zero. In accordance with
the scaling equation (\ref{eq:usKc}) we have
\begin{equation}
   \label{eq:uLKc}
   u_L(K_c) = u_\infty + A_uL^{-d+y_t} + \ldots,
\end{equation}
where $A_u$ is the amplitude of the leading finite-size
term in the internal energy density of the
system at the critical point. Calculations using formula
(\ref{eq:uLKc}) for a $4\times4\times\infty$ cluster using the
values $u_\infty=0.990\,637(47)$ \cite{HP98,HP97}
and $y_t=1.5865(14)$ \cite{BST99} give $A_u\simeq1.19$.

Since $A_u\ne0$, equation (\ref{eq:uu}) now only reproduces an
approximate value of $K_c$ . Also equation (\ref{eq:uf}) is not
satisfied exactly. Calculations of the position of the minimum
of the function (\ref{eq:F}) for the triad $(2,3,4)$ yield the
estimate $K_c=0.233003$. Unfortunately, the accuracy of
this estimate ($5.1\%$) is significantly inferior to the
errors of the estimates obtained from the other renormalization
schemes (see Table~\ref{tab:3DI}).


\section{Conclusions}

Nontrivial models with known values of the critical
coupling have been used to study phenomenological
renormalization schemes which for the same subsystem
sizes give more accurate results than
those used so far. We established that for Ising models
the renormalization group equations using combinations
$L^{1-2d}\kappa_L^{(4)}/\chi_L^2$ and $L^{1-d}\kappa_L^{(2)}/\chi_L$,
give uniformly converging bilateral estimates of $K_c$ where
the errors (in the three-dimensional, most important case)
are less than half those obtained in the Nightingale method.

Another result of this study is related to the exact
results. We have shown that by making the amplitudes
of the finite-size corrections to the internal
energy vanish, exact values of the critical temperatures
can be determined for the infinite two-dimensional Ising
and Potts models from the solutions for their partially
finite subsystems. In addition, we have observed that for
the two-dimensional Ising model the temperature derivative
of the inverse correlation length has absolutely no
finite-size corrections. This means that it is also
possible to find the exact value of the thermal critical
exponent for this model from the finite-size equation.

Both the conclusions reached from these results have
practical applications but separate analysis is required
to develop them.


\section*{Acknowledgments}

The author thanks A.~A.~Belavin for useful discussions and
valuable comments. Thanks are also due to
J.~Wosiek for showing interest in this work. This work
was supported financially by the Russian Foundation
for Basic Research (project no.~99-02-16472).



\newpage

\begin{table}
\caption{
Estimates of $K_c$ for a two-dimensional square Ising
lattice; $K_c^{exact}=0.440\,686\ldots$}
\label{tab:2DI}
\begin{tabular}{clll}
\hline
eq.&$(2,3)$&$(3,4)$&$(4,5)$\\[2mm]
\hline
 $(\kappa)$             &0.42236 &0.43088 $(-2.23\%)$ &0.43595\\
 $(\chi^{(4)}/\chi^2)$  &0.42593 &0.43242 $(-1.88\%)$ &0.43672\\
 $(\kappa^{(4)}/\chi^2)$&0.42596 &0.43243 $(-1.87\%)$ &0.43673\\[1mm]
 $(f^{s})$              &0.44324 &0.44168 $(+0.23\%)$ &0.44105\\
 $(\kappa^{(2)}/\chi)$  &0.47420 &0.45153 $(+2.64\%)$ &0.44626\\
\hline
\end{tabular}
\end{table}


\begin{table}
\caption{
Estimates of $K_c$ for a two-dimensional square three-site Potts
lattice; $K_c^{exact}=1.005\,052\ldots$}
\label{tab:2DP}
\begin{tabular}{clll}
\hline
eq.&$(2,3)$&$(3,4)$&$(4,5)$\\[2mm]
\hline
 $(\kappa)$               &0.96248 &0.98350 $(-2.1\%)$ &0.99467\\
 $(\kappa^{(1)^2}/\chi)$&0.99311 &0.99920 $(-0.6\%)$ &1.00380\\[1mm]
 $(f^{s})$                &1.00927 &1.00667 $(+0.2\%)$ &1.00565\\
\hline
\end{tabular}
\end{table}


\begin{table}
\caption{
Estimates  of $K_c$ for a three-dimensional simple
cubic Ising lattice;
$K_c^{exact}=0.221\,654\,59(10)$}
\label{tab:3DI}
\begin{tabular}{cll}
\hline
eq.&$(2,3)$&$(3,4)$\\[2mm]
\hline
 $(\kappa)$             &0.21340 &0.21826 $(-1.53\%)$\\
 $(\chi^{(4)}/\chi^2)$  &0.21823 &0.22002 $(-0.74\%)$\\
 $(\kappa^{(4)}/\chi^2)$&0.21824 &0.22006 $(-0.72\%)$\\[1mm]
 $(f^{s})$              &0.22354 &0.22236 $(+0.32\%)$\\
 $(\kappa^{(2)}/\chi)$  &0.22658 &0.22323 $(+0.71\%)$\\
\hline
\end{tabular}
\end{table}


\clearpage
\newpage

\begin{figure}
\topmargin 60mm

\setlength{\unitlength}{0.240900pt}
\ifx\plotpoint\undefined\newsavebox{\plotpoint}\fi
\sbox{\plotpoint}{\rule[-0.200pt]{0.400pt}{0.400pt}}%
\begin{picture}(1500,900)(0,0)
\font\gnuplot=cmr10 at 10pt
\gnuplot
\sbox{\plotpoint}{\rule[-0.200pt]{0.400pt}{0.400pt}}%
\put(161.0,163.0){\rule[-0.200pt]{4.818pt}{0.400pt}}
\put(141,163){\makebox(0,0)[r]{0}}
\put(1460.0,163.0){\rule[-0.200pt]{4.818pt}{0.400pt}}
\put(161.0,337.0){\rule[-0.200pt]{4.818pt}{0.400pt}}
\put(141,337){\makebox(0,0)[r]{5}}
\put(1460.0,337.0){\rule[-0.200pt]{4.818pt}{0.400pt}}
\put(161.0,511.0){\rule[-0.200pt]{4.818pt}{0.400pt}}
\put(141,511){\makebox(0,0)[r]{10}}
\put(1460.0,511.0){\rule[-0.200pt]{4.818pt}{0.400pt}}
\put(161.0,685.0){\rule[-0.200pt]{4.818pt}{0.400pt}}
\put(141,685){\makebox(0,0)[r]{15}}
\put(1460.0,685.0){\rule[-0.200pt]{4.818pt}{0.400pt}}
\put(161.0,859.0){\rule[-0.200pt]{4.818pt}{0.400pt}}
\put(141,859){\makebox(0,0)[r]{20}}
\put(1460.0,859.0){\rule[-0.200pt]{4.818pt}{0.400pt}}
\put(161.0,163.0){\rule[-0.200pt]{0.400pt}{4.818pt}}
\put(161,122){\makebox(0,0){0.1}}
\put(161.0,839.0){\rule[-0.200pt]{0.400pt}{4.818pt}}
\put(308.0,163.0){\rule[-0.200pt]{0.400pt}{4.818pt}}
\put(308,122){\makebox(0,0){}}
\put(308.0,839.0){\rule[-0.200pt]{0.400pt}{4.818pt}}
\put(454.0,163.0){\rule[-0.200pt]{0.400pt}{4.818pt}}
\put(454,122){\makebox(0,0){0.3}}
\put(454.0,839.0){\rule[-0.200pt]{0.400pt}{4.818pt}}
\put(601.0,163.0){\rule[-0.200pt]{0.400pt}{4.818pt}}
\put(601,122){\makebox(0,0){}}
\put(601.0,839.0){\rule[-0.200pt]{0.400pt}{4.818pt}}
\put(747.0,163.0){\rule[-0.200pt]{0.400pt}{4.818pt}}
\put(747,122){\makebox(0,0){0.5}}
\put(747.0,839.0){\rule[-0.200pt]{0.400pt}{4.818pt}}
\put(894.0,163.0){\rule[-0.200pt]{0.400pt}{4.818pt}}
\put(894,122){\makebox(0,0){}}
\put(894.0,839.0){\rule[-0.200pt]{0.400pt}{4.818pt}}
\put(1040.0,163.0){\rule[-0.200pt]{0.400pt}{4.818pt}}
\put(1040,122){\makebox(0,0){0.7}}
\put(1040.0,839.0){\rule[-0.200pt]{0.400pt}{4.818pt}}
\put(1187.0,163.0){\rule[-0.200pt]{0.400pt}{4.818pt}}
\put(1187,122){\makebox(0,0){}}
\put(1187.0,839.0){\rule[-0.200pt]{0.400pt}{4.818pt}}
\put(1333.0,163.0){\rule[-0.200pt]{0.400pt}{4.818pt}}
\put(1333,122){\makebox(0,0){0.9}}
\put(1333.0,839.0){\rule[-0.200pt]{0.400pt}{4.818pt}}
\put(1480.0,163.0){\rule[-0.200pt]{0.400pt}{4.818pt}}
\put(1480,122){\makebox(0,0){}}
\put(1480.0,839.0){\rule[-0.200pt]{0.400pt}{4.818pt}}
\put(161.0,163.0){\rule[-0.200pt]{317.747pt}{0.400pt}}
\put(1480.0,163.0){\rule[-0.200pt]{0.400pt}{167.666pt}}
\put(161.0,859.0){\rule[-0.200pt]{317.747pt}{0.400pt}}
\put(41,511){\makebox(0,0){$\cal F$}}
\put(820,61){\makebox(0,0){$K$}}
\put(425,415){\makebox(0,0)[l]{1}}
\put(483,320){\makebox(0,0)[l]{2}}
\put(1040,280){\makebox(0,0)[l]{3}}
\put(161.0,163.0){\rule[-0.200pt]{0.400pt}{167.666pt}}
\put(161,830){\usebox{\plotpoint}}
\multiput(161.58,822.61)(0.494,-2.139){27}{\rule{0.119pt}{1.780pt}}
\multiput(160.17,826.31)(15.000,-59.306){2}{\rule{0.400pt}{0.890pt}}
\multiput(176.58,760.30)(0.494,-1.929){25}{\rule{0.119pt}{1.614pt}}
\multiput(175.17,763.65)(14.000,-49.649){2}{\rule{0.400pt}{0.807pt}}
\multiput(190.58,708.60)(0.494,-1.523){27}{\rule{0.119pt}{1.300pt}}
\multiput(189.17,711.30)(15.000,-42.302){2}{\rule{0.400pt}{0.650pt}}
\multiput(205.58,664.38)(0.494,-1.284){27}{\rule{0.119pt}{1.113pt}}
\multiput(204.17,666.69)(15.000,-35.689){2}{\rule{0.400pt}{0.557pt}}
\multiput(220.58,626.67)(0.494,-1.195){25}{\rule{0.119pt}{1.043pt}}
\multiput(219.17,628.84)(14.000,-30.835){2}{\rule{0.400pt}{0.521pt}}
\multiput(234.58,594.37)(0.494,-0.976){27}{\rule{0.119pt}{0.873pt}}
\multiput(233.17,596.19)(15.000,-27.187){2}{\rule{0.400pt}{0.437pt}}
\multiput(249.58,565.71)(0.494,-0.873){27}{\rule{0.119pt}{0.793pt}}
\multiput(248.17,567.35)(15.000,-24.353){2}{\rule{0.400pt}{0.397pt}}
\multiput(264.58,539.98)(0.494,-0.791){25}{\rule{0.119pt}{0.729pt}}
\multiput(263.17,541.49)(14.000,-20.488){2}{\rule{0.400pt}{0.364pt}}
\multiput(278.58,518.37)(0.494,-0.668){27}{\rule{0.119pt}{0.633pt}}
\multiput(277.17,519.69)(15.000,-18.685){2}{\rule{0.400pt}{0.317pt}}
\multiput(293.58,498.59)(0.494,-0.600){27}{\rule{0.119pt}{0.580pt}}
\multiput(292.17,499.80)(15.000,-16.796){2}{\rule{0.400pt}{0.290pt}}
\multiput(308.58,480.69)(0.494,-0.570){25}{\rule{0.119pt}{0.557pt}}
\multiput(307.17,481.84)(14.000,-14.844){2}{\rule{0.400pt}{0.279pt}}
\multiput(322.00,465.92)(0.497,-0.494){27}{\rule{0.500pt}{0.119pt}}
\multiput(322.00,466.17)(13.962,-15.000){2}{\rule{0.250pt}{0.400pt}}
\multiput(337.00,450.92)(0.576,-0.493){23}{\rule{0.562pt}{0.119pt}}
\multiput(337.00,451.17)(13.834,-13.000){2}{\rule{0.281pt}{0.400pt}}
\multiput(352.00,437.92)(0.637,-0.492){19}{\rule{0.609pt}{0.118pt}}
\multiput(352.00,438.17)(12.736,-11.000){2}{\rule{0.305pt}{0.400pt}}
\multiput(366.00,426.92)(0.684,-0.492){19}{\rule{0.645pt}{0.118pt}}
\multiput(366.00,427.17)(13.660,-11.000){2}{\rule{0.323pt}{0.400pt}}
\multiput(381.00,415.92)(0.704,-0.491){17}{\rule{0.660pt}{0.118pt}}
\multiput(381.00,416.17)(12.630,-10.000){2}{\rule{0.330pt}{0.400pt}}
\multiput(395.00,405.93)(0.956,-0.488){13}{\rule{0.850pt}{0.117pt}}
\multiput(395.00,406.17)(13.236,-8.000){2}{\rule{0.425pt}{0.400pt}}
\multiput(410.00,397.93)(0.956,-0.488){13}{\rule{0.850pt}{0.117pt}}
\multiput(410.00,398.17)(13.236,-8.000){2}{\rule{0.425pt}{0.400pt}}
\multiput(425.00,389.93)(1.026,-0.485){11}{\rule{0.900pt}{0.117pt}}
\multiput(425.00,390.17)(12.132,-7.000){2}{\rule{0.450pt}{0.400pt}}
\multiput(439.00,382.93)(1.103,-0.485){11}{\rule{0.957pt}{0.117pt}}
\multiput(439.00,383.17)(13.013,-7.000){2}{\rule{0.479pt}{0.400pt}}
\multiput(454.00,375.93)(1.304,-0.482){9}{\rule{1.100pt}{0.116pt}}
\multiput(454.00,376.17)(12.717,-6.000){2}{\rule{0.550pt}{0.400pt}}
\multiput(469.00,369.93)(1.489,-0.477){7}{\rule{1.220pt}{0.115pt}}
\multiput(469.00,370.17)(11.468,-5.000){2}{\rule{0.610pt}{0.400pt}}
\multiput(483.00,364.94)(2.090,-0.468){5}{\rule{1.600pt}{0.113pt}}
\multiput(483.00,365.17)(11.679,-4.000){2}{\rule{0.800pt}{0.400pt}}
\multiput(498.00,360.94)(2.090,-0.468){5}{\rule{1.600pt}{0.113pt}}
\multiput(498.00,361.17)(11.679,-4.000){2}{\rule{0.800pt}{0.400pt}}
\multiput(513.00,356.94)(1.943,-0.468){5}{\rule{1.500pt}{0.113pt}}
\multiput(513.00,357.17)(10.887,-4.000){2}{\rule{0.750pt}{0.400pt}}
\multiput(527.00,352.95)(3.141,-0.447){3}{\rule{2.100pt}{0.108pt}}
\multiput(527.00,353.17)(10.641,-3.000){2}{\rule{1.050pt}{0.400pt}}
\multiput(542.00,349.95)(3.141,-0.447){3}{\rule{2.100pt}{0.108pt}}
\multiput(542.00,350.17)(10.641,-3.000){2}{\rule{1.050pt}{0.400pt}}
\put(557,346.17){\rule{2.900pt}{0.400pt}}
\multiput(557.00,347.17)(7.981,-2.000){2}{\rule{1.450pt}{0.400pt}}
\put(571,344.17){\rule{3.100pt}{0.400pt}}
\multiput(571.00,345.17)(8.566,-2.000){2}{\rule{1.550pt}{0.400pt}}
\put(586,342.17){\rule{3.100pt}{0.400pt}}
\multiput(586.00,343.17)(8.566,-2.000){2}{\rule{1.550pt}{0.400pt}}
\put(601,340.67){\rule{3.373pt}{0.400pt}}
\multiput(601.00,341.17)(7.000,-1.000){2}{\rule{1.686pt}{0.400pt}}
\put(615,339.67){\rule{3.614pt}{0.400pt}}
\multiput(615.00,340.17)(7.500,-1.000){2}{\rule{1.807pt}{0.400pt}}
\put(689,339.67){\rule{3.373pt}{0.400pt}}
\multiput(689.00,339.17)(7.000,1.000){2}{\rule{1.686pt}{0.400pt}}
\put(703,340.67){\rule{3.614pt}{0.400pt}}
\multiput(703.00,340.17)(7.500,1.000){2}{\rule{1.807pt}{0.400pt}}
\put(718,341.67){\rule{3.614pt}{0.400pt}}
\multiput(718.00,341.17)(7.500,1.000){2}{\rule{1.807pt}{0.400pt}}
\put(733,343.17){\rule{2.900pt}{0.400pt}}
\multiput(733.00,342.17)(7.981,2.000){2}{\rule{1.450pt}{0.400pt}}
\put(747,344.67){\rule{3.614pt}{0.400pt}}
\multiput(747.00,344.17)(7.500,1.000){2}{\rule{1.807pt}{0.400pt}}
\put(762,346.17){\rule{3.100pt}{0.400pt}}
\multiput(762.00,345.17)(8.566,2.000){2}{\rule{1.550pt}{0.400pt}}
\put(777,348.17){\rule{2.900pt}{0.400pt}}
\multiput(777.00,347.17)(7.981,2.000){2}{\rule{1.450pt}{0.400pt}}
\multiput(791.00,350.61)(3.141,0.447){3}{\rule{2.100pt}{0.108pt}}
\multiput(791.00,349.17)(10.641,3.000){2}{\rule{1.050pt}{0.400pt}}
\put(806,353.17){\rule{3.100pt}{0.400pt}}
\multiput(806.00,352.17)(8.566,2.000){2}{\rule{1.550pt}{0.400pt}}
\multiput(821.00,355.61)(2.918,0.447){3}{\rule{1.967pt}{0.108pt}}
\multiput(821.00,354.17)(9.918,3.000){2}{\rule{0.983pt}{0.400pt}}
\multiput(835.00,358.61)(3.141,0.447){3}{\rule{2.100pt}{0.108pt}}
\multiput(835.00,357.17)(10.641,3.000){2}{\rule{1.050pt}{0.400pt}}
\multiput(850.00,361.61)(2.918,0.447){3}{\rule{1.967pt}{0.108pt}}
\multiput(850.00,360.17)(9.918,3.000){2}{\rule{0.983pt}{0.400pt}}
\multiput(864.00,364.61)(3.141,0.447){3}{\rule{2.100pt}{0.108pt}}
\multiput(864.00,363.17)(10.641,3.000){2}{\rule{1.050pt}{0.400pt}}
\multiput(879.00,367.60)(2.090,0.468){5}{\rule{1.600pt}{0.113pt}}
\multiput(879.00,366.17)(11.679,4.000){2}{\rule{0.800pt}{0.400pt}}
\multiput(894.00,371.60)(1.943,0.468){5}{\rule{1.500pt}{0.113pt}}
\multiput(894.00,370.17)(10.887,4.000){2}{\rule{0.750pt}{0.400pt}}
\multiput(908.00,375.61)(3.141,0.447){3}{\rule{2.100pt}{0.108pt}}
\multiput(908.00,374.17)(10.641,3.000){2}{\rule{1.050pt}{0.400pt}}
\multiput(923.00,378.59)(1.601,0.477){7}{\rule{1.300pt}{0.115pt}}
\multiput(923.00,377.17)(12.302,5.000){2}{\rule{0.650pt}{0.400pt}}
\multiput(938.00,383.60)(1.943,0.468){5}{\rule{1.500pt}{0.113pt}}
\multiput(938.00,382.17)(10.887,4.000){2}{\rule{0.750pt}{0.400pt}}
\multiput(952.00,387.60)(2.090,0.468){5}{\rule{1.600pt}{0.113pt}}
\multiput(952.00,386.17)(11.679,4.000){2}{\rule{0.800pt}{0.400pt}}
\multiput(967.00,391.59)(1.601,0.477){7}{\rule{1.300pt}{0.115pt}}
\multiput(967.00,390.17)(12.302,5.000){2}{\rule{0.650pt}{0.400pt}}
\multiput(982.00,396.60)(1.943,0.468){5}{\rule{1.500pt}{0.113pt}}
\multiput(982.00,395.17)(10.887,4.000){2}{\rule{0.750pt}{0.400pt}}
\multiput(996.00,400.59)(1.601,0.477){7}{\rule{1.300pt}{0.115pt}}
\multiput(996.00,399.17)(12.302,5.000){2}{\rule{0.650pt}{0.400pt}}
\multiput(1011.00,405.59)(1.601,0.477){7}{\rule{1.300pt}{0.115pt}}
\multiput(1011.00,404.17)(12.302,5.000){2}{\rule{0.650pt}{0.400pt}}
\multiput(1026.00,410.59)(1.489,0.477){7}{\rule{1.220pt}{0.115pt}}
\multiput(1026.00,409.17)(11.468,5.000){2}{\rule{0.610pt}{0.400pt}}
\multiput(1040.00,415.59)(1.601,0.477){7}{\rule{1.300pt}{0.115pt}}
\multiput(1040.00,414.17)(12.302,5.000){2}{\rule{0.650pt}{0.400pt}}
\multiput(1055.00,420.59)(1.304,0.482){9}{\rule{1.100pt}{0.116pt}}
\multiput(1055.00,419.17)(12.717,6.000){2}{\rule{0.550pt}{0.400pt}}
\multiput(1070.00,426.59)(1.214,0.482){9}{\rule{1.033pt}{0.116pt}}
\multiput(1070.00,425.17)(11.855,6.000){2}{\rule{0.517pt}{0.400pt}}
\multiput(1084.00,432.59)(1.601,0.477){7}{\rule{1.300pt}{0.115pt}}
\multiput(1084.00,431.17)(12.302,5.000){2}{\rule{0.650pt}{0.400pt}}
\multiput(1099.00,437.59)(1.304,0.482){9}{\rule{1.100pt}{0.116pt}}
\multiput(1099.00,436.17)(12.717,6.000){2}{\rule{0.550pt}{0.400pt}}
\multiput(1114.00,443.59)(1.214,0.482){9}{\rule{1.033pt}{0.116pt}}
\multiput(1114.00,442.17)(11.855,6.000){2}{\rule{0.517pt}{0.400pt}}
\multiput(1128.00,449.59)(1.304,0.482){9}{\rule{1.100pt}{0.116pt}}
\multiput(1128.00,448.17)(12.717,6.000){2}{\rule{0.550pt}{0.400pt}}
\multiput(1143.00,455.59)(1.103,0.485){11}{\rule{0.957pt}{0.117pt}}
\multiput(1143.00,454.17)(13.013,7.000){2}{\rule{0.479pt}{0.400pt}}
\multiput(1158.00,462.59)(1.214,0.482){9}{\rule{1.033pt}{0.116pt}}
\multiput(1158.00,461.17)(11.855,6.000){2}{\rule{0.517pt}{0.400pt}}
\multiput(1172.00,468.59)(1.103,0.485){11}{\rule{0.957pt}{0.117pt}}
\multiput(1172.00,467.17)(13.013,7.000){2}{\rule{0.479pt}{0.400pt}}
\multiput(1187.00,475.59)(1.103,0.485){11}{\rule{0.957pt}{0.117pt}}
\multiput(1187.00,474.17)(13.013,7.000){2}{\rule{0.479pt}{0.400pt}}
\multiput(1202.00,482.59)(1.026,0.485){11}{\rule{0.900pt}{0.117pt}}
\multiput(1202.00,481.17)(12.132,7.000){2}{\rule{0.450pt}{0.400pt}}
\multiput(1216.00,489.59)(1.103,0.485){11}{\rule{0.957pt}{0.117pt}}
\multiput(1216.00,488.17)(13.013,7.000){2}{\rule{0.479pt}{0.400pt}}
\multiput(1231.00,496.59)(1.103,0.485){11}{\rule{0.957pt}{0.117pt}}
\multiput(1231.00,495.17)(13.013,7.000){2}{\rule{0.479pt}{0.400pt}}
\multiput(1246.00,503.59)(0.890,0.488){13}{\rule{0.800pt}{0.117pt}}
\multiput(1246.00,502.17)(12.340,8.000){2}{\rule{0.400pt}{0.400pt}}
\multiput(1260.00,511.59)(0.956,0.488){13}{\rule{0.850pt}{0.117pt}}
\multiput(1260.00,510.17)(13.236,8.000){2}{\rule{0.425pt}{0.400pt}}
\multiput(1275.00,519.59)(1.026,0.485){11}{\rule{0.900pt}{0.117pt}}
\multiput(1275.00,518.17)(12.132,7.000){2}{\rule{0.450pt}{0.400pt}}
\multiput(1289.00,526.59)(0.956,0.488){13}{\rule{0.850pt}{0.117pt}}
\multiput(1289.00,525.17)(13.236,8.000){2}{\rule{0.425pt}{0.400pt}}
\multiput(1304.00,534.59)(0.844,0.489){15}{\rule{0.767pt}{0.118pt}}
\multiput(1304.00,533.17)(13.409,9.000){2}{\rule{0.383pt}{0.400pt}}
\multiput(1319.00,543.59)(0.890,0.488){13}{\rule{0.800pt}{0.117pt}}
\multiput(1319.00,542.17)(12.340,8.000){2}{\rule{0.400pt}{0.400pt}}
\multiput(1333.00,551.59)(0.956,0.488){13}{\rule{0.850pt}{0.117pt}}
\multiput(1333.00,550.17)(13.236,8.000){2}{\rule{0.425pt}{0.400pt}}
\multiput(1348.00,559.59)(0.844,0.489){15}{\rule{0.767pt}{0.118pt}}
\multiput(1348.00,558.17)(13.409,9.000){2}{\rule{0.383pt}{0.400pt}}
\multiput(1363.00,568.59)(0.786,0.489){15}{\rule{0.722pt}{0.118pt}}
\multiput(1363.00,567.17)(12.501,9.000){2}{\rule{0.361pt}{0.400pt}}
\multiput(1377.00,577.59)(0.844,0.489){15}{\rule{0.767pt}{0.118pt}}
\multiput(1377.00,576.17)(13.409,9.000){2}{\rule{0.383pt}{0.400pt}}
\multiput(1392.00,586.59)(0.844,0.489){15}{\rule{0.767pt}{0.118pt}}
\multiput(1392.00,585.17)(13.409,9.000){2}{\rule{0.383pt}{0.400pt}}
\multiput(1407.00,595.58)(0.704,0.491){17}{\rule{0.660pt}{0.118pt}}
\multiput(1407.00,594.17)(12.630,10.000){2}{\rule{0.330pt}{0.400pt}}
\multiput(1421.00,605.59)(0.844,0.489){15}{\rule{0.767pt}{0.118pt}}
\multiput(1421.00,604.17)(13.409,9.000){2}{\rule{0.383pt}{0.400pt}}
\multiput(1436.00,614.58)(0.756,0.491){17}{\rule{0.700pt}{0.118pt}}
\multiput(1436.00,613.17)(13.547,10.000){2}{\rule{0.350pt}{0.400pt}}
\multiput(1451.00,624.58)(0.704,0.491){17}{\rule{0.660pt}{0.118pt}}
\multiput(1451.00,623.17)(12.630,10.000){2}{\rule{0.330pt}{0.400pt}}
\multiput(1465.00,634.58)(0.684,0.492){19}{\rule{0.645pt}{0.118pt}}
\multiput(1465.00,633.17)(13.660,11.000){2}{\rule{0.323pt}{0.400pt}}
\put(630.0,340.0){\rule[-0.200pt]{14.213pt}{0.400pt}}
\put(161,670){\usebox{\plotpoint}}
\multiput(161.58,664.05)(0.494,-1.694){27}{\rule{0.119pt}{1.433pt}}
\multiput(160.17,667.03)(15.000,-47.025){2}{\rule{0.400pt}{0.717pt}}
\multiput(176.58,614.72)(0.494,-1.488){25}{\rule{0.119pt}{1.271pt}}
\multiput(175.17,617.36)(14.000,-38.361){2}{\rule{0.400pt}{0.636pt}}
\multiput(190.58,574.71)(0.494,-1.181){27}{\rule{0.119pt}{1.033pt}}
\multiput(189.17,576.86)(15.000,-32.855){2}{\rule{0.400pt}{0.517pt}}
\multiput(205.58,540.26)(0.494,-1.010){27}{\rule{0.119pt}{0.900pt}}
\multiput(204.17,542.13)(15.000,-28.132){2}{\rule{0.400pt}{0.450pt}}
\multiput(220.58,510.50)(0.494,-0.938){25}{\rule{0.119pt}{0.843pt}}
\multiput(219.17,512.25)(14.000,-24.251){2}{\rule{0.400pt}{0.421pt}}
\multiput(234.58,485.04)(0.494,-0.771){27}{\rule{0.119pt}{0.713pt}}
\multiput(233.17,486.52)(15.000,-21.519){2}{\rule{0.400pt}{0.357pt}}
\multiput(249.58,462.26)(0.494,-0.702){27}{\rule{0.119pt}{0.660pt}}
\multiput(248.17,463.63)(15.000,-19.630){2}{\rule{0.400pt}{0.330pt}}
\multiput(264.58,441.45)(0.494,-0.644){25}{\rule{0.119pt}{0.614pt}}
\multiput(263.17,442.73)(14.000,-16.725){2}{\rule{0.400pt}{0.307pt}}
\multiput(278.58,423.81)(0.494,-0.531){27}{\rule{0.119pt}{0.527pt}}
\multiput(277.17,424.91)(15.000,-14.907){2}{\rule{0.400pt}{0.263pt}}
\multiput(293.00,408.92)(0.497,-0.494){27}{\rule{0.500pt}{0.119pt}}
\multiput(293.00,409.17)(13.962,-15.000){2}{\rule{0.250pt}{0.400pt}}
\multiput(308.00,393.92)(0.536,-0.493){23}{\rule{0.531pt}{0.119pt}}
\multiput(308.00,394.17)(12.898,-13.000){2}{\rule{0.265pt}{0.400pt}}
\multiput(322.00,380.92)(0.625,-0.492){21}{\rule{0.600pt}{0.119pt}}
\multiput(322.00,381.17)(13.755,-12.000){2}{\rule{0.300pt}{0.400pt}}
\multiput(337.00,368.92)(0.684,-0.492){19}{\rule{0.645pt}{0.118pt}}
\multiput(337.00,369.17)(13.660,-11.000){2}{\rule{0.323pt}{0.400pt}}
\multiput(352.00,357.92)(0.704,-0.491){17}{\rule{0.660pt}{0.118pt}}
\multiput(352.00,358.17)(12.630,-10.000){2}{\rule{0.330pt}{0.400pt}}
\multiput(366.00,347.93)(0.844,-0.489){15}{\rule{0.767pt}{0.118pt}}
\multiput(366.00,348.17)(13.409,-9.000){2}{\rule{0.383pt}{0.400pt}}
\multiput(381.00,338.93)(0.786,-0.489){15}{\rule{0.722pt}{0.118pt}}
\multiput(381.00,339.17)(12.501,-9.000){2}{\rule{0.361pt}{0.400pt}}
\multiput(395.00,329.93)(1.103,-0.485){11}{\rule{0.957pt}{0.117pt}}
\multiput(395.00,330.17)(13.013,-7.000){2}{\rule{0.479pt}{0.400pt}}
\multiput(410.00,322.93)(1.103,-0.485){11}{\rule{0.957pt}{0.117pt}}
\multiput(410.00,323.17)(13.013,-7.000){2}{\rule{0.479pt}{0.400pt}}
\multiput(425.00,315.93)(1.026,-0.485){11}{\rule{0.900pt}{0.117pt}}
\multiput(425.00,316.17)(12.132,-7.000){2}{\rule{0.450pt}{0.400pt}}
\multiput(439.00,308.93)(1.304,-0.482){9}{\rule{1.100pt}{0.116pt}}
\multiput(439.00,309.17)(12.717,-6.000){2}{\rule{0.550pt}{0.400pt}}
\multiput(454.00,302.93)(1.601,-0.477){7}{\rule{1.300pt}{0.115pt}}
\multiput(454.00,303.17)(12.302,-5.000){2}{\rule{0.650pt}{0.400pt}}
\multiput(469.00,297.93)(1.489,-0.477){7}{\rule{1.220pt}{0.115pt}}
\multiput(469.00,298.17)(11.468,-5.000){2}{\rule{0.610pt}{0.400pt}}
\multiput(483.00,292.94)(2.090,-0.468){5}{\rule{1.600pt}{0.113pt}}
\multiput(483.00,293.17)(11.679,-4.000){2}{\rule{0.800pt}{0.400pt}}
\multiput(498.00,288.94)(2.090,-0.468){5}{\rule{1.600pt}{0.113pt}}
\multiput(498.00,289.17)(11.679,-4.000){2}{\rule{0.800pt}{0.400pt}}
\multiput(513.00,284.95)(2.918,-0.447){3}{\rule{1.967pt}{0.108pt}}
\multiput(513.00,285.17)(9.918,-3.000){2}{\rule{0.983pt}{0.400pt}}
\multiput(527.00,281.95)(3.141,-0.447){3}{\rule{2.100pt}{0.108pt}}
\multiput(527.00,282.17)(10.641,-3.000){2}{\rule{1.050pt}{0.400pt}}
\multiput(542.00,278.95)(3.141,-0.447){3}{\rule{2.100pt}{0.108pt}}
\multiput(542.00,279.17)(10.641,-3.000){2}{\rule{1.050pt}{0.400pt}}
\put(557,275.17){\rule{2.900pt}{0.400pt}}
\multiput(557.00,276.17)(7.981,-2.000){2}{\rule{1.450pt}{0.400pt}}
\put(571,273.17){\rule{3.100pt}{0.400pt}}
\multiput(571.00,274.17)(8.566,-2.000){2}{\rule{1.550pt}{0.400pt}}
\put(586,271.67){\rule{3.614pt}{0.400pt}}
\multiput(586.00,272.17)(7.500,-1.000){2}{\rule{1.807pt}{0.400pt}}
\put(601,270.67){\rule{3.373pt}{0.400pt}}
\multiput(601.00,271.17)(7.000,-1.000){2}{\rule{1.686pt}{0.400pt}}
\put(615,269.67){\rule{3.614pt}{0.400pt}}
\multiput(615.00,270.17)(7.500,-1.000){2}{\rule{1.807pt}{0.400pt}}
\put(630,268.67){\rule{3.614pt}{0.400pt}}
\multiput(630.00,269.17)(7.500,-1.000){2}{\rule{1.807pt}{0.400pt}}
\put(674,268.67){\rule{3.614pt}{0.400pt}}
\multiput(674.00,268.17)(7.500,1.000){2}{\rule{1.807pt}{0.400pt}}
\put(645.0,269.0){\rule[-0.200pt]{6.986pt}{0.400pt}}
\put(703,269.67){\rule{3.614pt}{0.400pt}}
\multiput(703.00,269.17)(7.500,1.000){2}{\rule{1.807pt}{0.400pt}}
\put(718,270.67){\rule{3.614pt}{0.400pt}}
\multiput(718.00,270.17)(7.500,1.000){2}{\rule{1.807pt}{0.400pt}}
\put(733,272.17){\rule{2.900pt}{0.400pt}}
\multiput(733.00,271.17)(7.981,2.000){2}{\rule{1.450pt}{0.400pt}}
\put(747,274.17){\rule{3.100pt}{0.400pt}}
\multiput(747.00,273.17)(8.566,2.000){2}{\rule{1.550pt}{0.400pt}}
\put(762,275.67){\rule{3.614pt}{0.400pt}}
\multiput(762.00,275.17)(7.500,1.000){2}{\rule{1.807pt}{0.400pt}}
\put(777,277.17){\rule{2.900pt}{0.400pt}}
\multiput(777.00,276.17)(7.981,2.000){2}{\rule{1.450pt}{0.400pt}}
\multiput(791.00,279.61)(3.141,0.447){3}{\rule{2.100pt}{0.108pt}}
\multiput(791.00,278.17)(10.641,3.000){2}{\rule{1.050pt}{0.400pt}}
\put(806,282.17){\rule{3.100pt}{0.400pt}}
\multiput(806.00,281.17)(8.566,2.000){2}{\rule{1.550pt}{0.400pt}}
\multiput(821.00,284.61)(2.918,0.447){3}{\rule{1.967pt}{0.108pt}}
\multiput(821.00,283.17)(9.918,3.000){2}{\rule{0.983pt}{0.400pt}}
\put(835,287.17){\rule{3.100pt}{0.400pt}}
\multiput(835.00,286.17)(8.566,2.000){2}{\rule{1.550pt}{0.400pt}}
\multiput(850.00,289.61)(2.918,0.447){3}{\rule{1.967pt}{0.108pt}}
\multiput(850.00,288.17)(9.918,3.000){2}{\rule{0.983pt}{0.400pt}}
\multiput(864.00,292.61)(3.141,0.447){3}{\rule{2.100pt}{0.108pt}}
\multiput(864.00,291.17)(10.641,3.000){2}{\rule{1.050pt}{0.400pt}}
\multiput(879.00,295.60)(2.090,0.468){5}{\rule{1.600pt}{0.113pt}}
\multiput(879.00,294.17)(11.679,4.000){2}{\rule{0.800pt}{0.400pt}}
\multiput(894.00,299.61)(2.918,0.447){3}{\rule{1.967pt}{0.108pt}}
\multiput(894.00,298.17)(9.918,3.000){2}{\rule{0.983pt}{0.400pt}}
\multiput(908.00,302.60)(2.090,0.468){5}{\rule{1.600pt}{0.113pt}}
\multiput(908.00,301.17)(11.679,4.000){2}{\rule{0.800pt}{0.400pt}}
\multiput(923.00,306.61)(3.141,0.447){3}{\rule{2.100pt}{0.108pt}}
\multiput(923.00,305.17)(10.641,3.000){2}{\rule{1.050pt}{0.400pt}}
\multiput(938.00,309.60)(1.943,0.468){5}{\rule{1.500pt}{0.113pt}}
\multiput(938.00,308.17)(10.887,4.000){2}{\rule{0.750pt}{0.400pt}}
\multiput(952.00,313.60)(2.090,0.468){5}{\rule{1.600pt}{0.113pt}}
\multiput(952.00,312.17)(11.679,4.000){2}{\rule{0.800pt}{0.400pt}}
\multiput(967.00,317.60)(2.090,0.468){5}{\rule{1.600pt}{0.113pt}}
\multiput(967.00,316.17)(11.679,4.000){2}{\rule{0.800pt}{0.400pt}}
\multiput(982.00,321.60)(1.943,0.468){5}{\rule{1.500pt}{0.113pt}}
\multiput(982.00,320.17)(10.887,4.000){2}{\rule{0.750pt}{0.400pt}}
\multiput(996.00,325.60)(2.090,0.468){5}{\rule{1.600pt}{0.113pt}}
\multiput(996.00,324.17)(11.679,4.000){2}{\rule{0.800pt}{0.400pt}}
\multiput(1011.00,329.59)(1.601,0.477){7}{\rule{1.300pt}{0.115pt}}
\multiput(1011.00,328.17)(12.302,5.000){2}{\rule{0.650pt}{0.400pt}}
\multiput(1026.00,334.60)(1.943,0.468){5}{\rule{1.500pt}{0.113pt}}
\multiput(1026.00,333.17)(10.887,4.000){2}{\rule{0.750pt}{0.400pt}}
\multiput(1040.00,338.59)(1.601,0.477){7}{\rule{1.300pt}{0.115pt}}
\multiput(1040.00,337.17)(12.302,5.000){2}{\rule{0.650pt}{0.400pt}}
\multiput(1055.00,343.60)(2.090,0.468){5}{\rule{1.600pt}{0.113pt}}
\multiput(1055.00,342.17)(11.679,4.000){2}{\rule{0.800pt}{0.400pt}}
\multiput(1070.00,347.59)(1.489,0.477){7}{\rule{1.220pt}{0.115pt}}
\multiput(1070.00,346.17)(11.468,5.000){2}{\rule{0.610pt}{0.400pt}}
\multiput(1084.00,352.59)(1.601,0.477){7}{\rule{1.300pt}{0.115pt}}
\multiput(1084.00,351.17)(12.302,5.000){2}{\rule{0.650pt}{0.400pt}}
\multiput(1099.00,357.59)(1.601,0.477){7}{\rule{1.300pt}{0.115pt}}
\multiput(1099.00,356.17)(12.302,5.000){2}{\rule{0.650pt}{0.400pt}}
\multiput(1114.00,362.59)(1.489,0.477){7}{\rule{1.220pt}{0.115pt}}
\multiput(1114.00,361.17)(11.468,5.000){2}{\rule{0.610pt}{0.400pt}}
\multiput(1128.00,367.59)(1.601,0.477){7}{\rule{1.300pt}{0.115pt}}
\multiput(1128.00,366.17)(12.302,5.000){2}{\rule{0.650pt}{0.400pt}}
\multiput(1143.00,372.59)(1.304,0.482){9}{\rule{1.100pt}{0.116pt}}
\multiput(1143.00,371.17)(12.717,6.000){2}{\rule{0.550pt}{0.400pt}}
\multiput(1158.00,378.59)(1.489,0.477){7}{\rule{1.220pt}{0.115pt}}
\multiput(1158.00,377.17)(11.468,5.000){2}{\rule{0.610pt}{0.400pt}}
\multiput(1172.00,383.59)(1.304,0.482){9}{\rule{1.100pt}{0.116pt}}
\multiput(1172.00,382.17)(12.717,6.000){2}{\rule{0.550pt}{0.400pt}}
\multiput(1187.00,389.59)(1.601,0.477){7}{\rule{1.300pt}{0.115pt}}
\multiput(1187.00,388.17)(12.302,5.000){2}{\rule{0.650pt}{0.400pt}}
\multiput(1202.00,394.59)(1.214,0.482){9}{\rule{1.033pt}{0.116pt}}
\multiput(1202.00,393.17)(11.855,6.000){2}{\rule{0.517pt}{0.400pt}}
\multiput(1216.00,400.59)(1.304,0.482){9}{\rule{1.100pt}{0.116pt}}
\multiput(1216.00,399.17)(12.717,6.000){2}{\rule{0.550pt}{0.400pt}}
\multiput(1231.00,406.59)(1.304,0.482){9}{\rule{1.100pt}{0.116pt}}
\multiput(1231.00,405.17)(12.717,6.000){2}{\rule{0.550pt}{0.400pt}}
\multiput(1246.00,412.59)(1.214,0.482){9}{\rule{1.033pt}{0.116pt}}
\multiput(1246.00,411.17)(11.855,6.000){2}{\rule{0.517pt}{0.400pt}}
\multiput(1260.00,418.59)(1.304,0.482){9}{\rule{1.100pt}{0.116pt}}
\multiput(1260.00,417.17)(12.717,6.000){2}{\rule{0.550pt}{0.400pt}}
\multiput(1275.00,424.59)(1.026,0.485){11}{\rule{0.900pt}{0.117pt}}
\multiput(1275.00,423.17)(12.132,7.000){2}{\rule{0.450pt}{0.400pt}}
\multiput(1289.00,431.59)(1.304,0.482){9}{\rule{1.100pt}{0.116pt}}
\multiput(1289.00,430.17)(12.717,6.000){2}{\rule{0.550pt}{0.400pt}}
\multiput(1304.00,437.59)(1.103,0.485){11}{\rule{0.957pt}{0.117pt}}
\multiput(1304.00,436.17)(13.013,7.000){2}{\rule{0.479pt}{0.400pt}}
\multiput(1319.00,444.59)(1.026,0.485){11}{\rule{0.900pt}{0.117pt}}
\multiput(1319.00,443.17)(12.132,7.000){2}{\rule{0.450pt}{0.400pt}}
\multiput(1333.00,451.59)(1.304,0.482){9}{\rule{1.100pt}{0.116pt}}
\multiput(1333.00,450.17)(12.717,6.000){2}{\rule{0.550pt}{0.400pt}}
\multiput(1348.00,457.59)(1.103,0.485){11}{\rule{0.957pt}{0.117pt}}
\multiput(1348.00,456.17)(13.013,7.000){2}{\rule{0.479pt}{0.400pt}}
\multiput(1363.00,464.59)(0.890,0.488){13}{\rule{0.800pt}{0.117pt}}
\multiput(1363.00,463.17)(12.340,8.000){2}{\rule{0.400pt}{0.400pt}}
\multiput(1377.00,472.59)(1.103,0.485){11}{\rule{0.957pt}{0.117pt}}
\multiput(1377.00,471.17)(13.013,7.000){2}{\rule{0.479pt}{0.400pt}}
\multiput(1392.00,479.59)(1.103,0.485){11}{\rule{0.957pt}{0.117pt}}
\multiput(1392.00,478.17)(13.013,7.000){2}{\rule{0.479pt}{0.400pt}}
\multiput(1407.00,486.59)(0.890,0.488){13}{\rule{0.800pt}{0.117pt}}
\multiput(1407.00,485.17)(12.340,8.000){2}{\rule{0.400pt}{0.400pt}}
\multiput(1421.00,494.59)(1.103,0.485){11}{\rule{0.957pt}{0.117pt}}
\multiput(1421.00,493.17)(13.013,7.000){2}{\rule{0.479pt}{0.400pt}}
\multiput(1436.00,501.59)(0.956,0.488){13}{\rule{0.850pt}{0.117pt}}
\multiput(1436.00,500.17)(13.236,8.000){2}{\rule{0.425pt}{0.400pt}}
\multiput(1451.00,509.59)(0.890,0.488){13}{\rule{0.800pt}{0.117pt}}
\multiput(1451.00,508.17)(12.340,8.000){2}{\rule{0.400pt}{0.400pt}}
\multiput(1465.00,517.59)(0.956,0.488){13}{\rule{0.850pt}{0.117pt}}
\multiput(1465.00,516.17)(13.236,8.000){2}{\rule{0.425pt}{0.400pt}}
\put(689.0,270.0){\rule[-0.200pt]{3.373pt}{0.400pt}}
\sbox{\plotpoint}{\rule[-0.400pt]{0.800pt}{0.800pt}}%
\put(161,606){\usebox{\plotpoint}}
\multiput(162.41,595.43)(0.508,-1.519){23}{\rule{0.122pt}{2.547pt}}
\multiput(159.34,600.71)(15.000,-38.714){2}{\rule{0.800pt}{1.273pt}}
\multiput(177.41,552.39)(0.509,-1.370){21}{\rule{0.123pt}{2.314pt}}
\multiput(174.34,557.20)(14.000,-32.197){2}{\rule{0.800pt}{1.157pt}}
\multiput(191.41,517.09)(0.508,-1.095){23}{\rule{0.122pt}{1.907pt}}
\multiput(188.34,521.04)(15.000,-28.043){2}{\rule{0.800pt}{0.953pt}}
\multiput(206.41,486.19)(0.508,-0.918){23}{\rule{0.122pt}{1.640pt}}
\multiput(203.34,489.60)(15.000,-23.596){2}{\rule{0.800pt}{0.820pt}}
\multiput(221.41,459.48)(0.509,-0.874){21}{\rule{0.123pt}{1.571pt}}
\multiput(218.34,462.74)(14.000,-20.738){2}{\rule{0.800pt}{0.786pt}}
\multiput(235.41,436.74)(0.508,-0.670){23}{\rule{0.122pt}{1.267pt}}
\multiput(232.34,439.37)(15.000,-17.371){2}{\rule{0.800pt}{0.633pt}}
\multiput(250.41,416.96)(0.508,-0.635){23}{\rule{0.122pt}{1.213pt}}
\multiput(247.34,419.48)(15.000,-16.482){2}{\rule{0.800pt}{0.607pt}}
\multiput(265.41,398.37)(0.509,-0.569){21}{\rule{0.123pt}{1.114pt}}
\multiput(262.34,400.69)(14.000,-13.687){2}{\rule{0.800pt}{0.557pt}}
\multiput(278.00,385.09)(0.493,-0.508){23}{\rule{1.000pt}{0.122pt}}
\multiput(278.00,385.34)(12.924,-15.000){2}{\rule{0.500pt}{0.800pt}}
\multiput(293.00,370.08)(0.574,-0.509){19}{\rule{1.123pt}{0.123pt}}
\multiput(293.00,370.34)(12.669,-13.000){2}{\rule{0.562pt}{0.800pt}}
\multiput(308.00,357.08)(0.581,-0.511){17}{\rule{1.133pt}{0.123pt}}
\multiput(308.00,357.34)(11.648,-12.000){2}{\rule{0.567pt}{0.800pt}}
\multiput(322.00,345.08)(0.689,-0.512){15}{\rule{1.291pt}{0.123pt}}
\multiput(322.00,345.34)(12.321,-11.000){2}{\rule{0.645pt}{0.800pt}}
\multiput(337.00,334.08)(0.766,-0.514){13}{\rule{1.400pt}{0.124pt}}
\multiput(337.00,334.34)(12.094,-10.000){2}{\rule{0.700pt}{0.800pt}}
\multiput(352.00,324.08)(0.800,-0.516){11}{\rule{1.444pt}{0.124pt}}
\multiput(352.00,324.34)(11.002,-9.000){2}{\rule{0.722pt}{0.800pt}}
\multiput(366.00,315.08)(0.993,-0.520){9}{\rule{1.700pt}{0.125pt}}
\multiput(366.00,315.34)(11.472,-8.000){2}{\rule{0.850pt}{0.800pt}}
\multiput(381.00,307.08)(0.920,-0.520){9}{\rule{1.600pt}{0.125pt}}
\multiput(381.00,307.34)(10.679,-8.000){2}{\rule{0.800pt}{0.800pt}}
\multiput(395.00,299.08)(1.176,-0.526){7}{\rule{1.914pt}{0.127pt}}
\multiput(395.00,299.34)(11.027,-7.000){2}{\rule{0.957pt}{0.800pt}}
\multiput(410.00,292.07)(1.467,-0.536){5}{\rule{2.200pt}{0.129pt}}
\multiput(410.00,292.34)(10.434,-6.000){2}{\rule{1.100pt}{0.800pt}}
\multiput(425.00,286.07)(1.355,-0.536){5}{\rule{2.067pt}{0.129pt}}
\multiput(425.00,286.34)(9.711,-6.000){2}{\rule{1.033pt}{0.800pt}}
\multiput(439.00,280.07)(1.467,-0.536){5}{\rule{2.200pt}{0.129pt}}
\multiput(439.00,280.34)(10.434,-6.000){2}{\rule{1.100pt}{0.800pt}}
\multiput(454.00,274.06)(2.104,-0.560){3}{\rule{2.600pt}{0.135pt}}
\multiput(454.00,274.34)(9.604,-5.000){2}{\rule{1.300pt}{0.800pt}}
\put(469,267.34){\rule{3.000pt}{0.800pt}}
\multiput(469.00,269.34)(7.773,-4.000){2}{\rule{1.500pt}{0.800pt}}
\put(483,263.34){\rule{3.200pt}{0.800pt}}
\multiput(483.00,265.34)(8.358,-4.000){2}{\rule{1.600pt}{0.800pt}}
\put(498,259.34){\rule{3.200pt}{0.800pt}}
\multiput(498.00,261.34)(8.358,-4.000){2}{\rule{1.600pt}{0.800pt}}
\put(513,255.84){\rule{3.373pt}{0.800pt}}
\multiput(513.00,257.34)(7.000,-3.000){2}{\rule{1.686pt}{0.800pt}}
\put(527,252.84){\rule{3.614pt}{0.800pt}}
\multiput(527.00,254.34)(7.500,-3.000){2}{\rule{1.807pt}{0.800pt}}
\put(542,249.84){\rule{3.614pt}{0.800pt}}
\multiput(542.00,251.34)(7.500,-3.000){2}{\rule{1.807pt}{0.800pt}}
\put(557,247.34){\rule{3.373pt}{0.800pt}}
\multiput(557.00,248.34)(7.000,-2.000){2}{\rule{1.686pt}{0.800pt}}
\put(571,245.34){\rule{3.614pt}{0.800pt}}
\multiput(571.00,246.34)(7.500,-2.000){2}{\rule{1.807pt}{0.800pt}}
\put(586,243.84){\rule{3.614pt}{0.800pt}}
\multiput(586.00,244.34)(7.500,-1.000){2}{\rule{1.807pt}{0.800pt}}
\put(601,242.34){\rule{3.373pt}{0.800pt}}
\multiput(601.00,243.34)(7.000,-2.000){2}{\rule{1.686pt}{0.800pt}}
\put(630,240.84){\rule{3.614pt}{0.800pt}}
\multiput(630.00,241.34)(7.500,-1.000){2}{\rule{1.807pt}{0.800pt}}
\put(615.0,243.0){\rule[-0.400pt]{3.613pt}{0.800pt}}
\put(674,240.84){\rule{3.614pt}{0.800pt}}
\multiput(674.00,240.34)(7.500,1.000){2}{\rule{1.807pt}{0.800pt}}
\put(645.0,242.0){\rule[-0.400pt]{6.986pt}{0.800pt}}
\put(703,241.84){\rule{3.614pt}{0.800pt}}
\multiput(703.00,241.34)(7.500,1.000){2}{\rule{1.807pt}{0.800pt}}
\put(718,242.84){\rule{3.614pt}{0.800pt}}
\multiput(718.00,242.34)(7.500,1.000){2}{\rule{1.807pt}{0.800pt}}
\put(733,244.34){\rule{3.373pt}{0.800pt}}
\multiput(733.00,243.34)(7.000,2.000){2}{\rule{1.686pt}{0.800pt}}
\put(747,246.34){\rule{3.614pt}{0.800pt}}
\multiput(747.00,245.34)(7.500,2.000){2}{\rule{1.807pt}{0.800pt}}
\put(762,247.84){\rule{3.614pt}{0.800pt}}
\multiput(762.00,247.34)(7.500,1.000){2}{\rule{1.807pt}{0.800pt}}
\put(777,249.34){\rule{3.373pt}{0.800pt}}
\multiput(777.00,248.34)(7.000,2.000){2}{\rule{1.686pt}{0.800pt}}
\put(791,251.84){\rule{3.614pt}{0.800pt}}
\multiput(791.00,250.34)(7.500,3.000){2}{\rule{1.807pt}{0.800pt}}
\put(806,254.34){\rule{3.614pt}{0.800pt}}
\multiput(806.00,253.34)(7.500,2.000){2}{\rule{1.807pt}{0.800pt}}
\put(821,256.84){\rule{3.373pt}{0.800pt}}
\multiput(821.00,255.34)(7.000,3.000){2}{\rule{1.686pt}{0.800pt}}
\put(835,259.34){\rule{3.614pt}{0.800pt}}
\multiput(835.00,258.34)(7.500,2.000){2}{\rule{1.807pt}{0.800pt}}
\put(850,261.84){\rule{3.373pt}{0.800pt}}
\multiput(850.00,260.34)(7.000,3.000){2}{\rule{1.686pt}{0.800pt}}
\put(864,264.84){\rule{3.614pt}{0.800pt}}
\multiput(864.00,263.34)(7.500,3.000){2}{\rule{1.807pt}{0.800pt}}
\put(879,267.84){\rule{3.614pt}{0.800pt}}
\multiput(879.00,266.34)(7.500,3.000){2}{\rule{1.807pt}{0.800pt}}
\put(894,270.84){\rule{3.373pt}{0.800pt}}
\multiput(894.00,269.34)(7.000,3.000){2}{\rule{1.686pt}{0.800pt}}
\put(908,273.84){\rule{3.614pt}{0.800pt}}
\multiput(908.00,272.34)(7.500,3.000){2}{\rule{1.807pt}{0.800pt}}
\put(923,277.34){\rule{3.200pt}{0.800pt}}
\multiput(923.00,275.34)(8.358,4.000){2}{\rule{1.600pt}{0.800pt}}
\put(938,280.84){\rule{3.373pt}{0.800pt}}
\multiput(938.00,279.34)(7.000,3.000){2}{\rule{1.686pt}{0.800pt}}
\put(952,284.34){\rule{3.200pt}{0.800pt}}
\multiput(952.00,282.34)(8.358,4.000){2}{\rule{1.600pt}{0.800pt}}
\put(967,288.34){\rule{3.200pt}{0.800pt}}
\multiput(967.00,286.34)(8.358,4.000){2}{\rule{1.600pt}{0.800pt}}
\put(982,291.84){\rule{3.373pt}{0.800pt}}
\multiput(982.00,290.34)(7.000,3.000){2}{\rule{1.686pt}{0.800pt}}
\put(996,295.34){\rule{3.200pt}{0.800pt}}
\multiput(996.00,293.34)(8.358,4.000){2}{\rule{1.600pt}{0.800pt}}
\put(1011,299.34){\rule{3.200pt}{0.800pt}}
\multiput(1011.00,297.34)(8.358,4.000){2}{\rule{1.600pt}{0.800pt}}
\put(1026,303.34){\rule{3.000pt}{0.800pt}}
\multiput(1026.00,301.34)(7.773,4.000){2}{\rule{1.500pt}{0.800pt}}
\put(1040,307.34){\rule{3.200pt}{0.800pt}}
\multiput(1040.00,305.34)(8.358,4.000){2}{\rule{1.600pt}{0.800pt}}
\multiput(1055.00,312.38)(2.104,0.560){3}{\rule{2.600pt}{0.135pt}}
\multiput(1055.00,309.34)(9.604,5.000){2}{\rule{1.300pt}{0.800pt}}
\put(1070,316.34){\rule{3.000pt}{0.800pt}}
\multiput(1070.00,314.34)(7.773,4.000){2}{\rule{1.500pt}{0.800pt}}
\put(1084,320.34){\rule{3.200pt}{0.800pt}}
\multiput(1084.00,318.34)(8.358,4.000){2}{\rule{1.600pt}{0.800pt}}
\multiput(1099.00,325.38)(2.104,0.560){3}{\rule{2.600pt}{0.135pt}}
\multiput(1099.00,322.34)(9.604,5.000){2}{\rule{1.300pt}{0.800pt}}
\multiput(1114.00,330.38)(1.936,0.560){3}{\rule{2.440pt}{0.135pt}}
\multiput(1114.00,327.34)(8.936,5.000){2}{\rule{1.220pt}{0.800pt}}
\put(1128,334.34){\rule{3.200pt}{0.800pt}}
\multiput(1128.00,332.34)(8.358,4.000){2}{\rule{1.600pt}{0.800pt}}
\multiput(1143.00,339.38)(2.104,0.560){3}{\rule{2.600pt}{0.135pt}}
\multiput(1143.00,336.34)(9.604,5.000){2}{\rule{1.300pt}{0.800pt}}
\multiput(1158.00,344.38)(1.936,0.560){3}{\rule{2.440pt}{0.135pt}}
\multiput(1158.00,341.34)(8.936,5.000){2}{\rule{1.220pt}{0.800pt}}
\multiput(1172.00,349.38)(2.104,0.560){3}{\rule{2.600pt}{0.135pt}}
\multiput(1172.00,346.34)(9.604,5.000){2}{\rule{1.300pt}{0.800pt}}
\multiput(1187.00,354.38)(2.104,0.560){3}{\rule{2.600pt}{0.135pt}}
\multiput(1187.00,351.34)(9.604,5.000){2}{\rule{1.300pt}{0.800pt}}
\multiput(1202.00,359.38)(1.936,0.560){3}{\rule{2.440pt}{0.135pt}}
\multiput(1202.00,356.34)(8.936,5.000){2}{\rule{1.220pt}{0.800pt}}
\multiput(1216.00,364.39)(1.467,0.536){5}{\rule{2.200pt}{0.129pt}}
\multiput(1216.00,361.34)(10.434,6.000){2}{\rule{1.100pt}{0.800pt}}
\multiput(1231.00,370.38)(2.104,0.560){3}{\rule{2.600pt}{0.135pt}}
\multiput(1231.00,367.34)(9.604,5.000){2}{\rule{1.300pt}{0.800pt}}
\multiput(1246.00,375.39)(1.355,0.536){5}{\rule{2.067pt}{0.129pt}}
\multiput(1246.00,372.34)(9.711,6.000){2}{\rule{1.033pt}{0.800pt}}
\multiput(1260.00,381.38)(2.104,0.560){3}{\rule{2.600pt}{0.135pt}}
\multiput(1260.00,378.34)(9.604,5.000){2}{\rule{1.300pt}{0.800pt}}
\multiput(1275.00,386.39)(1.355,0.536){5}{\rule{2.067pt}{0.129pt}}
\multiput(1275.00,383.34)(9.711,6.000){2}{\rule{1.033pt}{0.800pt}}
\multiput(1289.00,392.39)(1.467,0.536){5}{\rule{2.200pt}{0.129pt}}
\multiput(1289.00,389.34)(10.434,6.000){2}{\rule{1.100pt}{0.800pt}}
\multiput(1304.00,398.39)(1.467,0.536){5}{\rule{2.200pt}{0.129pt}}
\multiput(1304.00,395.34)(10.434,6.000){2}{\rule{1.100pt}{0.800pt}}
\multiput(1319.00,404.39)(1.355,0.536){5}{\rule{2.067pt}{0.129pt}}
\multiput(1319.00,401.34)(9.711,6.000){2}{\rule{1.033pt}{0.800pt}}
\multiput(1333.00,410.39)(1.467,0.536){5}{\rule{2.200pt}{0.129pt}}
\multiput(1333.00,407.34)(10.434,6.000){2}{\rule{1.100pt}{0.800pt}}
\multiput(1348.00,416.39)(1.467,0.536){5}{\rule{2.200pt}{0.129pt}}
\multiput(1348.00,413.34)(10.434,6.000){2}{\rule{1.100pt}{0.800pt}}
\multiput(1363.00,422.40)(1.088,0.526){7}{\rule{1.800pt}{0.127pt}}
\multiput(1363.00,419.34)(10.264,7.000){2}{\rule{0.900pt}{0.800pt}}
\multiput(1377.00,429.39)(1.467,0.536){5}{\rule{2.200pt}{0.129pt}}
\multiput(1377.00,426.34)(10.434,6.000){2}{\rule{1.100pt}{0.800pt}}
\multiput(1392.00,435.40)(1.176,0.526){7}{\rule{1.914pt}{0.127pt}}
\multiput(1392.00,432.34)(11.027,7.000){2}{\rule{0.957pt}{0.800pt}}
\multiput(1407.00,442.40)(1.088,0.526){7}{\rule{1.800pt}{0.127pt}}
\multiput(1407.00,439.34)(10.264,7.000){2}{\rule{0.900pt}{0.800pt}}
\multiput(1421.00,449.39)(1.467,0.536){5}{\rule{2.200pt}{0.129pt}}
\multiput(1421.00,446.34)(10.434,6.000){2}{\rule{1.100pt}{0.800pt}}
\multiput(1436.00,455.40)(1.176,0.526){7}{\rule{1.914pt}{0.127pt}}
\multiput(1436.00,452.34)(11.027,7.000){2}{\rule{0.957pt}{0.800pt}}
\multiput(1451.00,462.40)(0.920,0.520){9}{\rule{1.600pt}{0.125pt}}
\multiput(1451.00,459.34)(10.679,8.000){2}{\rule{0.800pt}{0.800pt}}
\multiput(1465.00,470.40)(1.176,0.526){7}{\rule{1.914pt}{0.127pt}}
\multiput(1465.00,467.34)(11.027,7.000){2}{\rule{0.957pt}{0.800pt}}
\put(689.0,243.0){\rule[-0.400pt]{3.373pt}{0.800pt}}
\end{picture}

\label{fig:F}
\caption{
Behavior of the function ${\cal F}(K)$ for triads of Ising strip
clusters $(1,2,3)$, $(2,3,4)$, and $(3,4,5)$ given by curves $1$, $2$,
and $3$, respectively. In all cases the minimum of the function
is positioned at the exact value
$K_c={1\over2}\ln(1 +\sqrt2)=0.440686\ldots$.}
\clearpage
\end{figure}
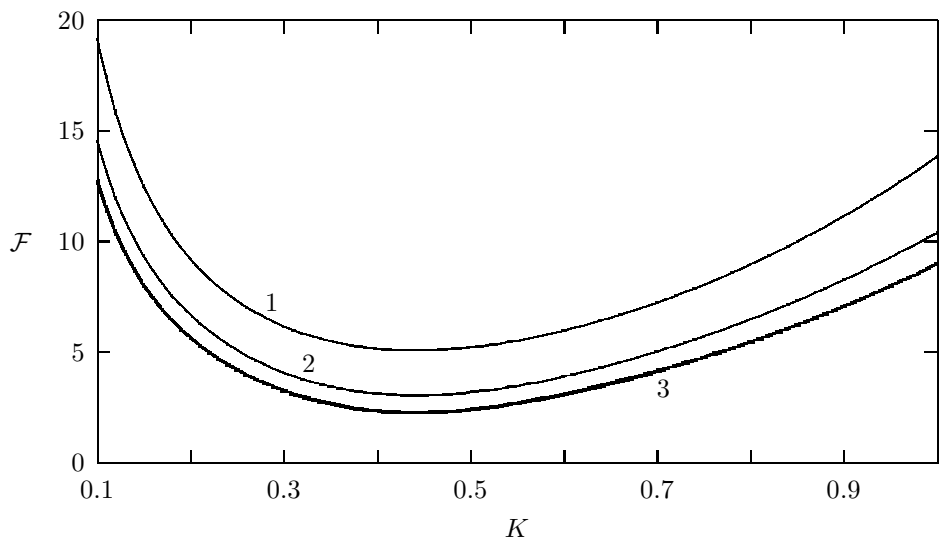



\newpage

\begin{thebibliography}{99}

\bibitem{M80} S.~Ma, {\em Modern Theory of Critical Phenomena\/},
(Benjamin, Reading, Mass., 1976; Mir, Moscow, 1980).

\bibitem{Nigh76} M.~P.~Nightingale, Physica A {\bf 83}, 561 (1976).

\bibitem{Nigh79} M.~P.~Nightingale, Proc. Kon. Ned. Akad. Wet. B
{\bf 82}, 235 (1979).

\bibitem{Nigh82} P.~Nightingale, J. Appl. Phys. {\bf 53}, 7927
(1982).

\bibitem{Nigh90} M.~P.~Nightingale, in {\em Finite Size Scaling and
Numerical Simulation of Statistical Systems\/}, ed. by V.~Privman,
World Scientific, Singapore (1990), p.~287.


\bibitem{DSS81} R.~R.~dos~Santos and L.~Sneddon, Phys. Rev. B {\bf 23},
3541 (1981).

\bibitem{B81a} K.~Binder, Phys. Rev. Lett. {\bf 47}, 693 (1981).

\bibitem{B81b} K.~Binder, Z. Phys. B {\bf 43}, 119 (1981).

\bibitem{I96} M.~Itakura, Preprint cond-mat/9611174.

\bibitem{F70} M.~Fisher, in {\em International School of "Enrico Fermi"
on Critical Phenomena, 1970\/}, Ed. by M. S. Green (Academic,
New York, 1971);
F.~J.~Dyson, E.~W.~Montroll, M.~Kac, and M.~Fisher, {\em Ustojchivost'
i Fazovye Perekhody\/} (Mir, Moscow, 1973), s.~245.

\bibitem{FB72} M.~E.~Fisher and M.~N.~Barber, Phys. Rev. Lett. {\bf 28},
1516 (1972).

\bibitem{B83} M.~N.~Barber, in {\em Phase Transitions and Critical
Phenomena\/}, ed. by C.~Domb and J.~L.~Lebowitz, Academic Press, London
(1983), Vol.~8, p.~145.

\bibitem{P90} V.Privman, in {\em Finite Size Scaling and Numerical
Simulation of Statistical Systems\/}, ed. by V.~Privman, World
Scientific, Singapore (1990), p.~1.

\bibitem{KS81} W.~Kinzel and M.~Schick, Phys. Rev. B {\bf 23}, 3435
(1981).

\bibitem{Yu00} M.~A.~Yurishchev, Nucl. Phys. B (Proc. Suppl.)
{\bf 83-84}, 727 (2000).

\bibitem{Yu94} M.~A.~Yurishchev, Phys. Rev. B {\bf 50}, 13\,533 (1994).

\bibitem{Yu97} M.~A.~Yurishchev, Phys. Rev. E {\bf 55}, 3915 (1997).

\bibitem{Bax85} R.~J.~Baxter, {\em Exactly Solved Models in Statistical
Mechanics\/} (Academic, New York, 1982; Mir, Moscow, 1985).

\bibitem{W82} F.~Y.~Wu, Rev. Mod. Phys. {\bf 54}, 235 (1982).

\bibitem{O44} L.~Onsager, Phys. Rev. {\bf 65}, 117 (1944).

\bibitem{SD85} H.~Saleur and B.~Derrida, J. Phys. (Paris) {\bf 46}, 1043
(1985).

\bibitem{K49} B.~Kaufman, Phys. Rev. {\bf 76}, 1232 (1949).

\bibitem{H66} K.~Huang, {\em Statistical Mechanics\/} (Wiley, New York,
1963; Mir, Moscow, 1966).

\bibitem{P52} R.~B.~Potts, Proc. Cambr. Phil. Soc. {\bf 48}, 106 (1952).

\bibitem{Bax73} R.~J.~Baxter, J. Phys. A {\bf 6}, L445 (1973).

\bibitem{Bax82} R.~J.~Baxter, Proc. Roy. Soc. London A {\bf 383}, 43
(1982).

\bibitem{W94} J.~Wosiek, Phys. Rev. B {\bf 49}, 15\,023 (1994).

\bibitem{BW94} Z.~Burda and J.~Wosiek, Nucl. Phys. B (Proc. Suppl.)
{\bf 34}, 677 (1994).

\bibitem{P95} A.~Pelizzola, Phys. Rev. B {\bf 51}, 12\,005 (1995).

\bibitem{PF84} V. Privman and M. E. Fisher, Phys. Rev. B {\bf 30}, 322
    (1984).

\bibitem{BST99} H.~W.~J.~Bl\"ote, L.~N.~Shchur, and A.~L.~Talapov,
Int. J. Mod. Phys. C {\bf 10}, 1137 (1999).

\bibitem{M89} K.~K.~Mon, Phys. Rev. B {\bf 39}, 467 (1989).

\bibitem{HP98} M.~Hasenbusch and K.~Pinn, Nucl. Phys. B (Proc. Suppl.)
{\bf 63}, 619 (1998).

\bibitem{HP97} M.~Hasenbusch and K.~Pinn, J. Phys. A {\bf 31}, 6157
(1998).

\end{thebibliography}
\end{document}